\documentstyle[psfrag,aps,preprint,epsfig,axodraw]{revtex}

\begin{document}

\psfrag{rho}{\hspace*{-3cm} $(a)~\rho_R = 1, ~a_1^R = 0$}
\psfrag{a}{\hspace*{-3cm}$(b)~\rho_R = 0.1,~a_1^R = m_0$}
\psfrag{mhalf}{$M_{1/2}~[\mbox{GeV}]$}
\psfrag{mzero}{$m_0~[\mbox{GeV}]$}

\tighten

\preprint{TU-677, DESY-02-221}

\title{Democratic (S)fermions and Lepton Flavor Violation}
\author{K. Hamaguchi$^{\rm a,b,}$\footnote{e-mail: koichi.hamaguchi@desy.de},
Mitsuru Kakizaki$^{\rm b,}$\footnote{e-mail: kakizaki@tuhep.phys.tohoku.ac.jp} 
and Masahiro Yamaguchi$^{\rm b,}$\footnote{e-mail: yama@tuhep.phys.tohoku.ac.jp}}
\address{$^{\rm a}$ Deutsches Elektronen-Synchrotron DESY, D-22603, Hamburg, Germany \\
and \\
$^{\rm b}$Department of Physics, Tohoku University,
Sendai 980-8578, Japan}
\date{\today}
\maketitle
\begin{abstract}
The democratic approach to account for  fermion masses and mixing is 
known to be successful not only in the quark sector but also in the
lepton sector. Here we extend this ansatz to supersymmetric standard
models, in which the K\"ahler potential obeys underlying $S_3$ flavor 
symmetries. The requirement of neutrino bi-large mixing angles constrains
the form of the K\"ahler potential for left-handed lepton multiplets. 
We find that right-handed sleptons can have non-degenerate masses and flavor
mixing, while left-handed sleptons are argued to have universal and hence
flavor-blind masses. This mass pattern is testable in future collider 
experiments when superparticle masses will be measured precisely. Lepton
flavor violation arises in this scenario. In particular, $\mu \to e \gamma$
is expected to be observed in a planning future experiment if supersymmetry
breaking scale is close to the weak scale. 
\end{abstract} 

\clearpage

\section{Introduction}

The flavor structure of quarks and leptons is one of the biggest puzzles of
particle physics. The minimal Standard Model has 19 unexplained parameters,
13 of which are for masses and mixing of the fermions. A few more parameters
(often associated with new particles) are needed to account for non-zero 
neutrino masses and their mixing. Some mechanism to explain  the fermion mass
structure is keenly desired.

The situation appears worse in supersymmetric (SUSY) extension of the
Standard Model.  The minimal SUSY Standard Model (MSSM) has additional
105 parameters, many of which are (SUSY breaking) masses and mixing of
scalar partners of quarks and leptons, {\it i.e.} sfermions.  Arbitrary 
mixing among the sfermions would cause too large flavor changing neutral current (FCNC)
far above the experimental bounds: this problem is referred to as the
SUSY flavor problem.

Thus the flavor problem in the SUSY Standard Model is two folds: 1)
the puzzle in the fermion mass structure and 2) the puzzle in the
sfermion mass structure.  How these two are linked with each other is
not clear.  In fact attempts have been made to separate these issues
and try to solve the latter independent of the former problem. This type
of approach  requires that the SUSY breaking occurring in a hidden
sector is mediated in a special way to our Standard Model sector. The 
SUSY flavor problem can be solved in this way, though at the same time
the resulting sfermion mass structure inherits no information on an
underlying theory of flavor which is presumably  a physics at very high
energy scale. 

Here we take an alternative approach to the flavor problem. Namely we
shall consider the case where  a mechanism which is responsible for the
fermion mass structure also dictates the sfermion mass structure 
\cite{Nir:1993mx,Dine:1993np,Pouliot:1993zm,Kaplan:1993ej,Pomarol:1995xc,Barbieri:1995uv,Hall:1995es}.
As was discussed earlier, such a strong connection between
the two aspects of the flavor problem in supersymmetry is not
mandatory. However this approach seems a natural one in the framework
of generic supergravity where one does not impose specific features on
the hidden sector and the messenger one.  Moreover, if this idea indeed
solves the SUSY flavor problem in nature, the structure of the
sfermion masses reflects the mechanism how the fermion masses are
generated, and thus one expects that future experiments will bring
some hints towards our understanding of the theory of flavor.

Explicitly in this paper we consider a democratic type of fermion mass
matrices (with diagonal breaking). The democratic ansatz 
\cite{Harari:1978yi,Koide:1983qe,Fritzsch:1995dj}
is very
attractive because it can beautifully explain bi-large mixing in the
neutrino sector while small mixing angles in the quark sector are
kept. Moreover the underlying symmetry is a permutation group of
three objects, $S_3$, which is one of the simplest groups and may be
naturally realized in various settings. A particularly interesting
possibility is that it can have a geometrical origin in brane-world
scenario \cite{Watari:2002fd}.

Here we assume that this democratic
principle can also operate in the sfermion masses. More precisely the
$S_3$ symmetry restricts possible forms of the K\"ahler potential, and
thus the scalar masses. Unlike the minimal supergravity, the resulting
sfermion masses are not degenerate among different generations, but they
are almost aligned with the fermion masses. A small violation of the
democracy will induce rather suppressed flavor violation in the
sfermion sector. Interestingly the effects to, e.g. $\mu \to e \gamma$
may be observed in near future experiments, as we will see later.

The paper is organized as follows. In the next section, we briefly
review some basic points of the democratic ansatz for fermion
masses. In Sec. 3, we extend this ansatz to the sfermion sector. We
postulate that the K\"ahler potential is invariant under the $S_3$
transformations among the generations.  This brings the characteristic
mass spectrum for squarks and sleptons as well as their flavor
mixing. The requirement that the neutrino sector has bi-large mixing
angles yields very non-trivial constraints on the K\"ahler potential. 
In Sec. 4, lepton flavor violation (LFV) is discussed. 
We show that it is plausible
that $\mu \to e \gamma$ will be observed in near future experiments in our
framework. In Sec. 5, we briefly discuss other flavor violation and
CP violation in quark sector.    The final section is devoted to 
conclusions and discussion.

\section{Democratic Fermion Masses}
In this section we briefly review the democratic ansatz for quarks and leptons
\cite{Fritzsch:1995dj}. We  write
the Yukawa interaction which generates the masses of quarks and charged 
leptons as follows:
\begin{eqnarray}
     W_Y&=& Q_L \cdot Y_q Q_R H 
\nonumber \\
        &=& Q_{Li} (Y_q)_{ ij} Q_{Rj} H,
\end{eqnarray}
where $Q_{Li}$ and $Q_{Rj}$ represent quarks or leptons of
(left-handed) SU(2)$_L$ doublets and (charge conjugated
right-handed) SU(2)$_L$ singlets, respectively, $H$ is a Higgs doublet
and $(Y_q)_{ij}$ represent Yukawa coupling constants. 
Indices $i$ and $j$ stand for
generations and run from 1 to 3. Here we have used the notion of
superpotential as well as chiral superfields for later convenience.

The basic idea of the democratic fermion is
to postulate that the Yukawa matrix $Y_q$
approximately takes a democratic form
\begin{eqnarray}
   Y_q \approx y_0^q J, \quad 
   J \equiv \frac{1}{3} 
      \left( 
      \begin{array}{ccc}
        1 & 1 & 1 \\
        1 & 1 & 1 \\
        1 & 1 & 1
      \end{array}
      \right),
\label{eq:democratic-mass}
\end{eqnarray}
with some parameter $y^q_0$. 
It follows immediately that the Yukawa matrix is invariant under $S_3(Q_L) \times
S_3(Q_R)$ transformation, where
$S_3(Q_L)$ represents a permutation group among the three generations
of $Q_{Li}$'s, and $S_3(Q_R)$ of $Q_{Ri}$'s. We call the basis which gives
(\ref{eq:democratic-mass}) a democratic basis.

It is sometimes convenient to change the basis by a matrix
\begin{eqnarray}
      A=\left(
          \begin{array}{ccc}
                1/\sqrt{2} & 1/\sqrt{6} & 1/\sqrt{3} \\
              -1/\sqrt{2}  & 1/\sqrt{6} & 1/\sqrt{3} \\
                 0  &  -2/\sqrt{6}      & 1/\sqrt{3} 
           \end{array}
         \right).
\end{eqnarray}
We call this basis a canonical basis. In this basis the three-dimensional
reducible representation $\underline{3}$ of $S_3$ is decomposed into a
two-dimensional irreducible representation $\underline{2}$ and a
one-dimensional (trivial) representation $\underline{1}$, namely
$\underline{3}=\underline{2} \oplus \underline{1}$. The above
mass matrix (\ref{eq:democratic-mass}) becomes proportional to
\begin{eqnarray}
   T \equiv    \left(
     \begin{array}{ccc}
      0 & 0 & 0 \\
      0 & 0 & 0 \\
      0 & 0 & 1
     \end{array}
    \right)
\end{eqnarray}
where only the field corresponding to $\underline{1}$ has non-zero mass, which
should be identified with the third generation in this democratic limit.
\footnote{$S_3$ flavor symmetry was also considered in Ref. \cite{Hall:1995es},
where the three generations transform as  $\underline{2}\oplus
\underline{1}_A$ with $\underline{1}_A$ being a non-trivial one dimensional
representation. The breaking pattern of the $S_3$ symmetry and hence the
mass matrix in Ref. \cite{Hall:1995es} is different from the democratic case 
we now  consider.}

To be more realistic, we need to include symmetry breaking terms. 
Let us first discuss the case of quarks.
Following Ref. \cite{Koide:1983qe}, we take in the democratic basis
\begin{eqnarray}
  Y_q= y_0^q \left[
      J
     +\left(
       \begin{array}{ccc}
        -\epsilon_q & 0 & 0 \\
            0   & \epsilon_q & 0 \\
            0   & 0  & \delta_q
       \end{array}
      \right)
      \right],
\label{eq:quarks_d}
\end{eqnarray}
where $q=u$ or $d$.
Here we have assumed $\epsilon_q \ll \delta_q \ll 1$. 
An advantage of the breaking terms in the form given above
becomes transparent if one moves to the canonical basis. There the first
$2 \times 2$ entries of the matrix are of the Fritzsch type
\cite{Fritzsch:1977vd}:
\begin{eqnarray}
  Y_q = y_0^q \left(
    \begin{array}{ccc}
      0 & -\sqrt{1/3} \epsilon_q & - \sqrt{2/3} \epsilon_q \\
      -\sqrt{1/3} \epsilon_q & (2/3) \delta_q & - (\sqrt{2}/3) \delta_q \\
       - \sqrt{2/3} \epsilon_q & - (\sqrt{2}/3) \delta_q & 1 + (1/3) \delta_q
    \end{array}
  \right),
\end{eqnarray}
from which one
immediately obtains a phenomenologically favorable relation for
the Cabibbo angle $\theta_C$:
\begin{eqnarray}
      \sin \theta_C \approx\sqrt{\frac{m_d}{m_s}} - \sqrt{\frac{m_u}{m_c}}
                   \approx \sqrt{\frac{m_d}{m_s}}
\end{eqnarray}
The mass matrix (\ref{eq:quarks_d}) 
is diagonalized by $U_q=A B_q$. The explicit
form of $B_q$ is found in the literature \cite{Koide:1983qe,Fritzsch:1995dj}.
Here it is sufficient to note that they are approximately
\begin{eqnarray}
        B_q=
       \left(
        \begin{array}{ccc}
         1 & O(\epsilon_q/\delta_q) & O(\epsilon_q) \\
        O(\epsilon_q/\delta_q) & 1 & O(\delta_q) \\
        O(\epsilon_q) & O(\delta_q) & 1 
       \end{array}
      \right)
\end{eqnarray}
and the ratio of the mass eigenvalues is roughly given by 
\begin{eqnarray}
      m^q_1 : m^q_2 : m^q_3 \approx 
    \frac{\epsilon_q^2}{\delta_q} : \delta_q : 1
\end{eqnarray}
A comparison with data determines the magnitudes of the breaking parameters. 
For instance, for the down-type quarks, we find
\begin{eqnarray}
     \delta_d \sim 10^{-1}, \quad \epsilon_d \sim 10^{-2}.
\end{eqnarray}
A similar expression can be obtained for the up-type quark sector, with much
smaller breaking parameters:
\begin{eqnarray}
  \delta_u \sim 10^{-2}, \quad \epsilon_u \sim 10^{-4}.
\end{eqnarray}
  The Cabibbo-Kobayashi-Maskawa (CKM) matrix can be obtained by 
\begin{eqnarray}
        U_{\mbox{CKM}}=\left(AB_u \right)^{\dagger} AB_d=B_u^{\dagger} B_d.
\end{eqnarray}
Thus the matrix $A$ cancels out, and the remaining one is  a matrix with  small mixing 
angles.  It has been argued that the CKM matrix obtained this way gives 
a phenomenologically successful pattern \cite{Koide:1983qe}.  
A detailed analysis of the
quark masses and the CKM matrix in the democratic ansatz will be renewed
elsewhere \cite{prep}.

Let us now turn to the lepton sector. The Yukawa matrix of the charged
leptons is assumed to be of the same form as those of the quarks
\cite{Fritzsch:1995dj}:
\begin{eqnarray}
  Y_l= y_0^l \left[
     J
     +\left(
       \begin{array}{ccc}
        -\epsilon_l & 0 & 0 \\
            0   & \epsilon_l & 0 \\
            0   & 0  & \delta_l
       \end{array}
      \right)
      \right].
\label{eq:quarks}
\end{eqnarray}
It possesses the flavor symmetry $S_3(L_L) \times S_3(E_R)$ where  
$S_3(L_L)$ and
$S_3(E_R)$ stand for permutation symmetries among the three generations of
$SU(2)_L$ doublet leptons and singlet leptons, respectively. The 
diagonalization can
be performed in exactly the same manner as the case of the quarks. 
The mass ratio among the three generations implies 
\begin{eqnarray} 
        \delta_l \sim 10^{-1}, \quad \epsilon_l \sim 10^{-2},
\end{eqnarray}
like in the case of the down-type quarks.

The situation is quite different in the neutrino mass matrix, which is given
by the bilinear of the doublet leptons. Thus it has two $S_3(L_L)$ invariants:
$J$ and $I$, the latter of  which denotes the $3\times 3$ unit matrix.
As was discussed by Ref. \cite{Fritzsch:1995dj}, the latter
should be taken to obtain bi-large mixing angles
which are strongly suggested 
by the atmospheric and the solar neutrino anomalies
\cite{Fukuda:1998mi,Fukuda:1998fd,Apollonio:1999ae}
\footnote{Very recently the first results from the KamLAND experiment
have been announced, which exclude all solutions to the solar neutrino 
problem but the large-angle matter-enhanced one 
\cite{KamLAND}.}.
This choice may be
justified in a brane-world scenario where each generation of the doublet 
leptons is localized on a different point in  extra dimensions 
\cite{Watari:2002fd}. 
With diagonal breaking terms, we have the following mass matrix
\begin{eqnarray}
  M_{\nu}=m_{\nu} \left[
         I
       +\left(
         \begin{array}{ccc}
           -\epsilon_{\nu} & 0 & 0 \\
           0 & \epsilon_{\nu} & 0 \\
           0  & 0  & \delta_{\nu}
         \end{array}
        \right)
       \right].
       \label{eq:numass}
\end{eqnarray}
We find the Maki-Nakagawa-Sakata (MNS)
matrix is approximated to be $A^{T}$, and thus
\begin{eqnarray}
       \sin^2 2 \theta_{\mbox{solar}} \approx 1, ~~~
       \sin^2 2 \theta_{\mbox{atm}} \approx 8/9, ~~~ U_{e3} \approx 0.
\end{eqnarray}

To summarize this section, the democratic ansatz with the diagonal breaking
works well not only in the quark sector but also in the lepton sector.

\section{Democratic sfermions}
In this section we extend this democratic principle 
to the soft SUSY breaking terms. 
To begin with, 
let us recall the SUSY flavor problem  in scenarios where 
supergravity interaction mediates the SUSY breaking:
Arbitrary couplings in the K\"ahler potential between the visible sector and
the hidden sector immediately 
cause disastrously large FCNC.
There are several proposals to this problem.
Some of the them attribute to separation 
between the two sectors by postulating special forms of the K\"ahler 
potential.
On the other hand, in the minimal supergravity and its variants \cite{Hall:iz}
a type of U$(n)$ flavor symmetry in the 
K\"ahler metric, with $n$ being the number of generations,
is implicitly assumed 
without any justification, while the Yukawa sector maximally breaks it.
Here we will show  that the K\"ahler potential controlled by the
$S_3$ symmetry solves the SUSY flavor problem, and at the same time
provides some interesting phenomenological consequences
on the sfermion masses and mixing,  
which can be tested by future experiments.

The $S_3$ symmetry allows the following K\"ahler potential for the MSSM fields 
$Q_i$:
\begin{eqnarray}
  K = [Z^Q_I I + Z^Q_J J]_{ij} Q^\dag_i Q_j, 
      \label{eq:kahler}
\end{eqnarray}
where $Z_I^Q$ and $Z_J^Q$ are, in general, functions of (hidden sector) fields.
In the following, we are interested in only their vacuum expectation values
which can be expanded in terms of the Grassmann variables $\theta$ and $\bar
\theta$:
\begin{eqnarray}
  Z^Q_{I,J} & = & z^Q_{I,J} + ( \theta^2 a^Q_{I,J} + h.c. )
  - \theta^2 \bar{\theta}^2 m_{QI,J}^2. 
\end{eqnarray}
Notice that $a^Q_{I,J}$ and $m_{QI,J}^2$ come from the SUSY breaking. 
The existence of the democratic part $Z^Q_J$ plays essential roles
in our arguments. In fact, the non-universal K\"ahler metric coming from 
$z_J^Q$ leads to non-universal kinetic terms, and hence affects the fermion
masses and their mixing. Furthermore the democratic part generates 
non-universal soft scalar masses, as we will explain in detail.

Using the matrix $A$ the K\"ahler metric for the fields $Q_i$ in 
eq. (\ref{eq:kahler}) is 
diagonalized as 
\begin{eqnarray}
   [z^Q_I I + z^Q_J T]_{ij}.
\end{eqnarray}
That is, we obtain the non-universal kinetic terms which are written explicitly
\begin{eqnarray}
  K \supset \left( 
      \begin{array}{ccc}
        z^Q_I & 0 & 0 \\
        0 & z^Q_I & 0 \\
        0 & 0 & z^Q_I + z^Q_J
      \end{array}
      \right)_{ij} Q^\dag_i Q_j.
\end{eqnarray}
After rescaling the fields using the diagonal matrix
\begin{eqnarray}
      C_Q & \equiv & 
    {\rm diag}\left(1/\sqrt{z^Q_I}, 1/\sqrt{z^Q_I}, 1/\sqrt{z^Q_I(1 + r_Q)} \right), \quad r_Q = z^Q_J/z^Q_I, 
\label{eq:rescale}
\end{eqnarray}
we obtain canonically normalized kinetic terms. In the presence of $z_J^Q$,
we call the field basis obtained this way the canonical basis.

Suppose that the Yukawa matrices
have the same structure as the ones discussed in the previous section:
\begin{eqnarray}
  Y & = & y_{0} \left[
    J
     +\left(
       \begin{array}{ccc}
        -\epsilon & 0 & 0 \\
            0   & \epsilon & 0 \\
            0   & 0  & \delta
       \end{array}
      \right)
      \right].
      \label{eq:yukawa_d}
\end{eqnarray}
Diagonalization of the kinetic terms by $A$ with subsequent rescaling 
transforms
the Yukawa matrix (\ref{eq:yukawa_d})
 into the one in the canonical basis, which is written 
\begin{eqnarray}
  Y  =  y_{0}  C_L \left[ T+ A^T
      \left(
      \begin{array}{ccc}
        -\epsilon & 0 & 0 \\
        0 & \epsilon & 0 \\
        0 & 0 & \delta
      \end{array}
      \right) A \right] C_R
    \label{eq:yukawa_c}
\end{eqnarray}
where $C_{L,R}$ denote the rescaling matrices (\ref{eq:rescale})
for $Q_{L,R}$.
Notice that in the canonical basis 
$Y$ (\ref{eq:yukawa_c}) is already almost diagonalized.

The presence of the non-universal K\"ahler metric modifies the masses and
mixing of the fermions.  Let us consider the diagonalization of 
the Yukawa matrix $Y$ (\ref{eq:yukawa_c}) given in the 
canonical basis:
\begin{eqnarray}
  U_L^T Y U_R & = & \mbox{diag}(y_1, y_2, y_3).
\end{eqnarray}
It is easy to show that 
the eigenvalues are approximately 
\begin{eqnarray}
    y_1 = Y_0 \left( \frac{\Delta}{3} - \frac{\Xi}{6} \right), \quad
    y_2 = Y_0 \left( \frac{\Delta}{3} + \frac{\Xi}{6} \right), \quad
    y_3 = Y_0 \left( 1 + \frac{\delta}{3} \right)
\end{eqnarray}
with
\begin{eqnarray}
  Y_{0} & = & \frac{y_{0}}{\sqrt{z_I^L z_I^R (1 + r_L)(1 + r_R)}},\quad 
  \Xi = 2 ( \Delta^2 + 3 E^2)^{1/2}, \nonumber \\
  \Delta & = & \sqrt{(1 + r_L)(1 + r_R)} \delta,\quad 
  E = \sqrt{(1 + r_L)(1 + r_R)} \epsilon.
\end{eqnarray}
Also the unitary matrices $U_L$ and $U_R$ which diagonalize the matrix are
found to be  
\begin{eqnarray}
  U_{L,R} =  \left(
    \begin{array}{ccc}
      \cos \theta & - \sin \theta & - \Lambda^{L,R} \sin 2 \theta \\
      \sin \theta & \cos \theta & - \Lambda^{L,R} \cos 2 \theta \\
      \Lambda^{L,R} \sin 3 \theta & \Lambda^{L,R} \cos 3 \theta & 1
    \end{array}
  \right),
\label{eq:ULR}
\end{eqnarray}
where 
\begin{eqnarray}
  \tan 2 \theta  = \frac{\sqrt{3} \epsilon}{\delta}, \quad 
  \Lambda^{L,R} = \sqrt{1 + r_{L,R}} \frac{\xi}{3 \sqrt{2}}, \quad
  \xi = 2 ( \delta^2 + 3 \epsilon^2)^{1/2}.
  \label{eq:lambda}
\end{eqnarray}
As far as the normalization factor $1+r_{L,R}$ is of order unity, the
masses and mixing angles of the quarks and charged leptons are not
drastically modified.  On the other hand, this factor is very important
for the neutrino mixing, as we will discuss later.

Let us now discuss the soft SUSY breaking terms. 
After the $A$ rotation, soft scalar mass terms for sfermions $\tilde{q}_i$
are written
\begin{eqnarray}
  - {\cal L} & \supset & (m^2_Q)_{ij} \tilde{q}^\dag_i \tilde{q}_j, 
  \nonumber \\
  m^2_Q & = & (m^2_{QI} I + m^2_{QJ} T)
  + (a^Q_I I + a^Q_J T)(z^Q_I I + z^Q_J T)^{-1}(a^Q_I I + a^Q_J T).
\end{eqnarray}
Since the second term of $m_Q^2$ is rewritten in terms of 
the linear combination of 
$I$ and $T$, hereafter we absorb this term into $m^2_{QI,J}$.
To go to the canonical basis, 
we need to multiply $C_Q$ from both sides, 
so that they become
\begin{eqnarray}
  m^2_Q & = & C_Q ( m^2_{QI} I + m^2_{QJ} T) C_Q \nonumber \\
  & = & m^2_{Q0} \left[ \left( 
    \begin{array}{ccc}
      1 & 0 & 0 \\
      0 & 1 & 0 \\
      0 & 0 & 1
    \end{array}
    \right) + \rho_Q \left( 
    \begin{array}{ccc}
      0 & 0 & 0 \\
      0 & 0 & 0 \\
      0 & 0 & 1
    \end{array}
    \right) \right], \quad
  m^2_{Q0} =  
  \frac{m^2_{QI}}{z_I^Q}, \quad 
  \rho_Q = \frac{z^Q_I m_{QJ}^2 - z_J^Q m^2_{QI}}{(z^Q_I + z^Q_J)m_{QI}^2}
  \label{eq:scalarmass}
\end{eqnarray}
We find that the quantity $\rho_Q$ becomes non-zero if 
the ratio of the democratic part to the unit part 
in the $D$-components of $Z$, $m^2_{QJ}/m^2_{QI}$, is different from 
the same ratio in the lowest components, $r_Q = z^Q_J/z^Q_I$. 
We expect that this is generically the case and thus $\rho_Q$ is of
order unity. The existence of the second term of the mass-squared matrix
leads to very interesting consequences. As one can readily see from 
eq. (\ref{eq:scalarmass}), the 
third generation sfermion has
a different mass than the degenerate 
first two generation masses, that is
\begin{eqnarray}
  m^2_{1st} = m^2_{2nd} \ne m^2_{3rd}.
\end{eqnarray}
Furthermore, in the fermion-mass basis where the fermion masses are diagonal,
there arise non-zero off diagonal elements in the sfermion mass-squared
matrix, leading to flavor mixing mediated by the sfermions. In fact, in
this basis, the mass-squared matrix becomes of the following form:
\begin{eqnarray}
  m^2_Q &=& m_{Q0}^2 \left[ \left(
      \begin{array}{ccc}
        1 & 0 & 0\\
        0 & 1 & 0 \\
        0 & 0 & 1
      \end{array}
    \right) + \rho_Q \left(
      \begin{array}{ccc}
        \Lambda^2 \sin^2 3 \theta & \Lambda^2 \sin 3 \theta \cos 3 \theta & 
        \Lambda \sin 3 \theta \\
        \Lambda^2 \sin 3 \theta \cos 3 \theta & \Lambda^2 \cos^2 3 \theta &
        \Lambda \cos 3 \theta \\
        \Lambda \sin 3 \theta  & \Lambda \cos 3 \theta & 1
      \end{array}
    \right) \right] \,, \nonumber \\
  & \sim & m^2_{Q0} \left(
    \begin{array}{ccc}
      1 & O(\rho \Delta E) & O(\rho E) \\
      O(\rho \Delta E) & 1 & O(\rho \Delta) \\
      O(\rho E) & O(\rho \Delta) & 1 + \rho
    \end{array} \right), \quad \rho \sim O(1),
\end{eqnarray}
where the coefficients in each entries can be computed by using 
eq.~(\ref{eq:ULR}) and $\Lambda$ here corresponds to $\Lambda_L$ or $\Lambda_R$
in eq. (\ref{eq:lambda})

Let us next turn to  scalar trilinear couplings, {\it i.e.}  $A$-terms.
Following the same procedure, we find that 
the $A$-terms in the canonical basis are given as
\begin{eqnarray}
  - {\cal L} \supset \sum_{Q} \frac{\partial W}{\partial \tilde{q}_i} 
  \left[(z_I^Q I + z_J^Q T)^{-1}(a_I^Q I + a_J^Q T)\right]_{ij} \tilde{q}_j.
\end{eqnarray}
As in the previous case, 
disagreement between $z^Q_J/z^Q_I$ and $a^Q_J/a^Q_I$ induces
small off-diagonal elements in the fermion-mass basis, resulting in
again the flavor mixing. An inspection  shows that the above equation
is written
\begin{eqnarray}
     \sum_{Q} \frac{\partial W}{\partial \tilde{q}_i} 
  \left[ a_0^Q I +a_1^Q T \right]_{ij} \tilde{q}_j,
  \label{eq:a-term}
\end{eqnarray}
where
\begin{eqnarray}
       a_0^Q=\frac{a_I^Q}{z_I^Q}, \quad 
       a_1^Q=\frac{a_J^Q z_I^Q-a_I^Q z_J^Q}{(z_I^Q + z_J^Q) z_I^Q}.
\end{eqnarray}
With the superpotential of the form
$ W=Q_L \cdot Y Q_R H $, 
eq. (\ref{eq:a-term}) reduces to 
\begin{eqnarray}
  (a_0^L +a_0^R +a_0^H) \tilde q_{L} \cdot Y \tilde q_R H
  + a_1^L \tilde q_L \cdot T Y \tilde q_R H 
  + a_1^R \tilde q_L \cdot Y T \tilde q_R H.
\end{eqnarray}
One can move to the fermion mass basis by multiplying $U_L^T$ and
$U_R$ from left and right, respectively. The matrix for the trilinear
couplings becomes of the following form
\begin{eqnarray}
  A_q & =  &     (a_0^L +a_0^R +a_0^H) 
    \left( \begin{array}{ccc}
           y_1 & 0 & 0 \\
           0  & y_2 & 0 \\
           0  & 0   & y_3 
          \end{array} \right)
    +   a_1^L \left( U_L^T  T U_L^* \right) 
          \left( \begin{array}{ccc}
           y_1 & 0 & 0 \\
           0  & y_2 & 0 \\
           0  & 0   & y_3 
          \end{array} \right) \nonumber \\ 
   & &  + a_1^R     \left( \begin{array}{ccc}
           y_1 & 0 & 0 \\
           0  & y_2 & 0 \\
           0  & 0   & y_3 
          \end{array} \right)
          \left( U_R^{\dagger} T U_R \right), 
  \label{eq:a}
\end{eqnarray}
and 
\begin{eqnarray}
    U_{L,R}^{\dagger} T U_{L,R} & = & 
\left(
    \begin{array}{ccc}
      (\Lambda^{L,R})^2 \sin^2 3 \theta & (\Lambda^{L,R})^2 \sin 3 \theta \cos 3 \theta & 
      (\Lambda^{L,R})^2 \sin 3 \theta \\
      (\Lambda^{L,R})^2 \sin 3 \theta \cos 3 \theta & (\Lambda^{L,R})^2 \cos^2 3 \theta &
      \Lambda^{L,R} \cos 3 \theta \\
      \Lambda^{L,R} \sin 3 \theta  & \Lambda^{L,R} \cos 3 \theta & 1
    \end{array}
  \right) \nonumber \\
 & \sim & 
  \left(\begin{array}{ccc}
          E^2 & \Delta E & E \\
          \Delta E & \Delta^2 & \Delta \\
           E   & \Delta & 1 
         \end{array}
    \right).
\end{eqnarray}
We find that the flavor mixing terms will not become symmetric matrices. This
will play an important role when discussing lepton flavor violation.

Following Ref. \cite{Gabbiani:1996hi}, we now introduce some notation on the
sfermion mass terms in the fermion mass basis
\begin{eqnarray}
  (m^2_{LL})_{ij} \tilde{q}_{Li}^* \tilde{q}_{Lj}
  + (m^2_{RR})_{ij} \tilde{q}_{Ri}^* \tilde{q}_{Rj}
  + (m^2_{LR})_{ij} \tilde{q}_{Li}^* \tilde{q}_{Rj}
  + (m^2_{RL})_{ij} \tilde{q}_{Ri}^* \tilde{q}_{Lj}, \quad
  m^{2\dag}_{LR} = m^2_{RL}
\end{eqnarray}
where each is given as
\begin{eqnarray}
  (m^2_{LL})_{ij} & = & (m^2_{Q_L})_{ij}, \quad 
  (m^2_{RR})_{ij} = (m^{2T}_{Q_R})_{ij}, \nonumber \\
  (m^2_{LR})_{ij} & = & (A_y^*)_{ij} \langle H \rangle, \quad
  (m^2_{RL})_{ij} = (A_y^T)_{ij} \langle H \rangle.
\end{eqnarray}
Here we have written only SUSY breaking masses explicitly, and suppressed SUSY
contributions.
For later use, it is convenient to define 
\begin{eqnarray}
    (\delta_{ij})_{LL} \equiv \frac{(m^2_{LL})_{ij}}{m^2}, \quad
    (\delta_{ij})_{RR} \equiv \frac{(m^2_{RR})_{ij}}{m^2}, \nonumber \\
    (\delta_{ij})_{LR} \equiv \frac{(m^2_{LR})_{ij}}{m^2}, \quad 
    (\delta_{ij})_{RL} \equiv \frac{(m^2_{RL})_{ij}}{m^2},
\end{eqnarray}
where $m$ represents a typical mass of the corresponding sfermion.

In this paper, we assume that the K\"ahler potential is 
invariant under the $S_3$ symmetry and do not include possible $S_3$ breaking
contributions.  This assumption is, however, not stable against radiative
corrections. In particular, renormalization group effects involving the
Yukawa couplings will violate the symmetry in the K\"ahler potential, which
generate different patterns of violations of mass degeneracy and flavor
mixing. We will return to this point in the subsequent sections.

\subsection{(S)lepton Sector}
In this subsection, we would like to focus on the lepton sector and
discuss the slepton masses in detail.

Following the general argument given above, we can perform the diagonalization
of the Yukawa matrix for the charged leptons. The results are
\begin{eqnarray}
    y_e = Y_0^l \left( \frac{\Delta_l}{3} - \frac{\Xi_l}{6} \right), \quad
    y_\mu = Y_0^l \left( \frac{\Delta_l}{3} + \frac{\Xi_l}{6} \right), \quad
    y_\tau = Y_0^l \left( 1 + \frac{\delta_l}{3} \right),
\end{eqnarray}
and 
\begin{eqnarray}
  U_{L,R} & = & \left(
    \begin{array}{ccc}
      \cos \theta_l & - \sin \theta_l & - \Lambda^{L,R} \sin 2 \theta_l \\
      \sin \theta_l & \cos \theta_l & - \Lambda^{L,R} \cos 2 \theta_l \\
      \Lambda^{L,R} \sin 3 \theta_l & \Lambda^{L,R} \cos 3 \theta_l & 1
    \end{array}
  \right).
\end{eqnarray}

Next we consider the neutrino mass matrix. We  assume that the neutrino 
mass matrix is the same as the one discussed in Section 2 
(eq. (\ref{eq:numass})).
It becomes, in the canonical basis, 
\begin{eqnarray}
  M_\nu & = & m_\nu C_L \left[ \left( 
      \begin{array}{ccc}
        1 & 0 & 0 \\
        0 & 1 & 0 \\
        0 & 0 & 1
      \end{array}
      \right) + A^T \left( 
      \begin{array}{ccc}
        0 & 0 & 0 \\
        0 & \epsilon_\nu & 0 \\
        0 & 0 & \delta_\nu
      \end{array}
      \right) A \right] C_L  
    \label{eq:nu_c}
\end{eqnarray}
Without $C_L$, the first term, {\it i.e.} the unit matrix, 
of eq.~(\ref{eq:nu_c})
is irrelevant to the diagonalization, and thus the mass matrix itself is
diagonalized by the matrix $A$. It is this fact which realizes the
bi-large mixing angles. However, when $r_L$ is non-zero and  $C_L$ is 
not  proportional to the unit matrix, this is no longer the case. This is 
because the first term in eq~(\ref{eq:nu_c}) is proportional to
$\mbox{diag}(1,1,1+r_L)$ and the diagonalizing matrix is different from $A$. In
fact, it is an easy task to see that the whole mass matrix in this case is
diagonalized by a matrix close to the unit matrix (with small angles). 
Combining the fact that the diagonalizing matrix for the charged lepton 
sector contains only small mixing angles, we find that 
mixing angles of the neutrinos would be all small
when the $r_L$ is sizably different from zero. 
On the other hand,
in the limit of $r_L \ll \epsilon_\nu, \delta_\nu$, 
the first term in (\ref{eq:nu_c}) can be regarded as a unit matrix.
Then the unitary matrix arises only from the second term in (\ref{eq:nu_c})
and equals to $A$, 
from which we can reproduce bi-large mixing angles in the neutrino sector.
We conclude that the absence of the democratic part in the K\"ahler metric 
$z_J^L$ of the doublet leptons is required by the neutrino phenomenology.

This observation strongly implies that the whole democratic part of the
K\"ahler potential for the doublet leptons, namely $Z_J^L=z_J^L
+(\theta^2 a_J^L +h.c.)-\theta^2 \bar \theta^2 m_{LJ}^2$, should be absent. 
Though the absence of only the lowest component, $z_J^L$, is sufficient for the
bi-large mixing to be realized,  we argue that
it is unnatural and thus  unlikely.  Assuming the absence of
the whole democratic part, namely $z_J^L=a_J^L=m^2_{LJ}=0$, we 
immediately conclude that 
all left-handed slepton masses are degenerate:
\begin{eqnarray}
  m^2_{\tilde e_L} = m^2_{\tilde \mu_L} = m^2_{\tilde \tau_L}, 
\label{eq:left-handed-sleptons}
\end{eqnarray}
at the energy scale where these soft scalar masses are given, presumably
at a scale close to the Planck scale. At the same time there will be no
flavor mixing in the left-handed slepton mass-squared matrix. 
On the other hand, the right-handed
lepton sector generically has the democratic part in its K\"ahler potential, 
and thus the right-handed stau has a different
mass from those in the first two generations:
\begin{eqnarray}
  m^2_{\tilde e_R}= m^2_{\tilde \mu_R} \neq m^2_{\tilde \tau_R}.
\label{eq:right-handed-sleptons}
\end{eqnarray}
Furthermore its presence also leads flavor mixing of the right-handed  
slepton mass-squared matrix of the slepton sector.
As for the $A$-terms, the absence of the democratic part for the left-handed 
leptons implies $a_1^L=0$, and thus the matrix for the $A$-terms (\ref{eq:a})
reduces to
\begin{eqnarray}
    A_l = (a_0^L +a_0^R +a_0^H)  \left(
    \begin{array}{ccc}
      y_e  & 0 & 0 \\
      0 & y_\mu & 0 \\
      0 & 0 & y_\tau
    \end{array}
  \right) 
+ a_1^R Y_0^l \left(
    \begin{array}{ccc}
      O(E^4/\Delta) &  O(E^3) & O(E^3/\Delta) \\
      O(\Delta^2 E) & O(\Delta^3) &  O(\Delta^2) \\
      O( E) & O(\Delta) & 1 
    \end{array} \right).
\label{eq:A-sleptons}
\end{eqnarray}
It is straightforward to compute numerical values of the matrix in 
the second term, once $\Delta_l$ and $E_l$
are fixed by the charged lepton masses
\footnote{We neglect $r_L$ since it is extremely small.}
\begin{eqnarray}
  \Delta_l \approx \frac{3 m_\mu}{2 m_\tau - m_\mu/\sqrt{1 + r_R}}, \quad 
  E_l \approx \sqrt{\frac{4 m_e}{3 m_\mu}}\,\Delta_l.
\label{eq:Delta-E}
\end{eqnarray}
Notice that they are insensitive to the choice of $r_R$ 
as far as $\sqrt{1 + r_R}$ is of order unity.

The mass spectrum in eqs.~(\ref{eq:left-handed-sleptons}) and
(\ref{eq:right-handed-sleptons}) is characteristic of the democratic
sfermion ansatz. In fact the degeneracy of the left-handed slepton
masses is strongly suggested by realizing the bi-large neutrino mixing
angles. Such a constraint does not come to the right-handed sector,
and thus the right-handed stau can have a different mass which is
consistent with the $S_3$ symmetry. This situation should be compared
with the renormalization group (RG) effect due to right-handed
neutrino Yukawa couplings, which leads the non-degeneracy in the
left-handed slepton masses \cite{Moroi:1993ui}. The non-degeneracy in
the right-handed sector can also be induced in SU(5) grand unified
theories (GUTs) by the RG effect above the GUT scale
\cite{Barbieri:1994pv}. One possible way to distinguish our case with
the GUT contribution is that in our case the right-handed stau can be
either light or heavy, while the RG flow always makes the right-handed stau
lighter than the other generations. 

\section{Lepton flavor violation}
We are now at a position to discuss lepton flavor violation. FCNC
processes in the quark sector will be briefly discussed in the
subsequent section.  In our setup, we will show that the predicted rate
of $\mu \to e \gamma$ is within reach of planning experiments as far as
the sparticle masses are of order 100 GeV.  We also
find that the rate of $\tau \to \mu \gamma$ is smaller than the present
experimental bound by at least two orders of magnitudes.

In SUSY models, superparticle exchange loop diagrams induce 
the dipole moment operators
\begin{eqnarray}
    {\cal L}_{\rm eff}  &= & e \frac{m_{l_j}}{2} \bar{l}_i \sigma_{\mu \nu} 
  F^{\mu \nu} (L_{ij} P_L + R_{ij} P_R) l_j
\label{eq:effective-op}
\end{eqnarray}
where $m_{l_j}$ denotes a charged lepton mass of generation $j$, 
$P_{L,R}=(1 \mp \gamma^5)/2$ indicate the chiral projection operators, and
$L_{ij}$ and $R_{ij}$ are coefficients dictated by the parameters in the 
SUSY models.
Flavor violating processes take place  if off-diagonal components in 
scalar  mass matrices exist.
For lepton flavor violation, the present experimental bound
\cite{Brooks:1999pu}
Br$(\mu \to e \gamma) < 1.2 \times 10^{-11}$ constrains the 
off-diagonal elements 
of the slepton masses. Here we give an estimate by using an approximate
formula similar to Ref. \cite{Gabbiani:1996hi}, where we only consider the 
diagrams given in Fig. \ref{fig:lfv}
with Bino and Wino loops. More details will be
found in Appendix. The constraints obtained in this approximation are  
\begin{eqnarray}
  (\delta^l_{12})_{LL} &\lesssim & ({\rm a~few}) \times 10^{-3} 
 \left( \frac{m_{\tilde{l}}}{100 ~\mbox{GeV}} \right)^2, \nonumber \\
  (\delta^l_{12})_{RR}  & \lesssim & ({\rm a~few})\times 10^{-3}
 \left( \frac{m_{\tilde{l}}}{100 ~\mbox{GeV}} \right)^2, \nonumber \\
  (\delta^l_{12})_{LR}, ~ (\delta^l_{12})_{RL} & \lesssim & ({\rm a~few}) \times 10^{-6}
 \left( \frac{m_{\tilde{l}}}{100 ~\mbox{GeV}} \right)^2,
\end{eqnarray}
where $m_{\tilde{l}}$ represents a characteristic mass scale of the sleptons, 
and
the constraints get weaker as $m_{\tilde{l}}$ increases. 
We note that  exact numbers are sensitive to
the ratios of the gaugino masses and the slepton masses, and the values
given above should be taken as typical ones.  

These constraints should be compared with the expected values of the
off-diagonal elements in the framework of the democratic sfermions.
For $m_{\tilde{l}} \sim 100$ GeV and the
scalar trilinear parameter  $a_1^R \sim m_{\tilde{l}}$, we find 
\begin{eqnarray}
  (\delta^l_{12})_{LL} & \sim &0, \quad 
  (\delta^l_{12})_{RR} \sim \rho_R \Delta_l E_l \sim 10^{-3}, 
  \nonumber \\
  (\delta^l_{12})_{LR} & \sim & 
  \frac{a_1^R \Delta_l E_l m_e}{m_{\tilde{l}}^2} 
  \sim 10^{-8}, \quad
  (\delta^l_{12})_{RL} \sim 
  \frac{a_1^R \Delta_l E_l m_\mu}{m_{\tilde{l}}^2} 
  \sim 10^{-6}.
\end{eqnarray}
The comparison shows that the flavor mixing expected in our framework is
already comparable with the experimental constraints if the SUSY breaking
scale is around the weak scale. Put another way,  we naturally expect that 
the $\mu \to e \gamma$ process will be seen in a near future
experiment if the sparticles in the loops have masses of order 100 GeV. 

The same comparison also indicates that the off-diagonal mixing
components $(\delta_{12})_{RR}$ and $(\delta_{12})_{RL}$ give
comparable and dominant contributions to $\mu \to e \gamma$. Both
contributions generate non-zero $L_{12}$ in
eq.~(\ref{eq:effective-op}), with vanishing $R_{12}$. It is
interesting to compare this with other scenarios of lepton flavor
violation. For instance, $(m^2_{LL})_{12}$ dominates when the RG
effect due to the right-handed neutrino Yukawa interactions gives
considerable flavor violation \cite{Borzumati:1986qx,Hisano:1995nq}. 
In this case, only
$R_{12}$ becomes non-zero. On the other hand, $(m^2_{RR})_{12}$ will
dominate in the minimal SU(5) GUT if the RG contributions above the
GUT scale are important \cite{Barbieri:1994pv}, which gives non-zero
$L_{12}$ term, as in our scenario.  It is interesting to note that the
case with non-zero $L_{12}$ may be distinguishable from the other case
with non-zero $R_{12}$ if polarized muon beam is available, by
measuring the angular distribution of the decaying particles
\cite{Kuno:1996kv}.

\begin{figure}[ht]
  \begin{center}
    \makebox[0cm]{
      \scalebox{1.0}{\hspace*{-3cm}
        \includegraphics{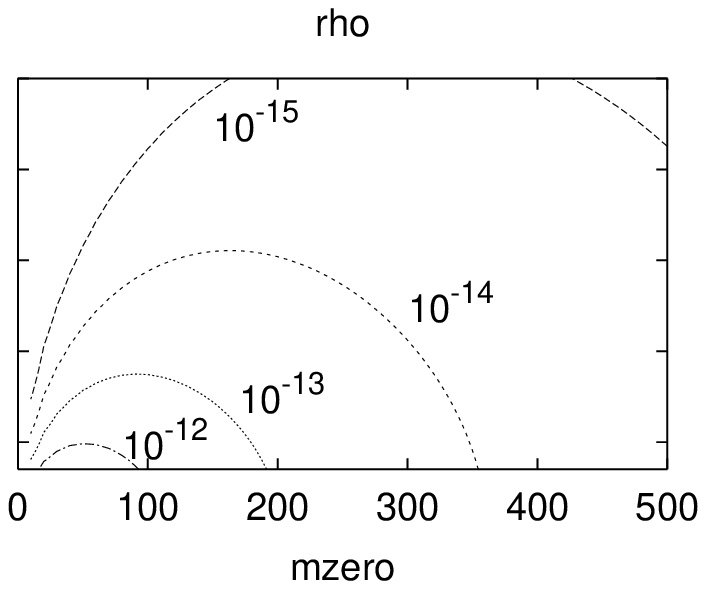} \hspace*{-5.5cm}
        \includegraphics{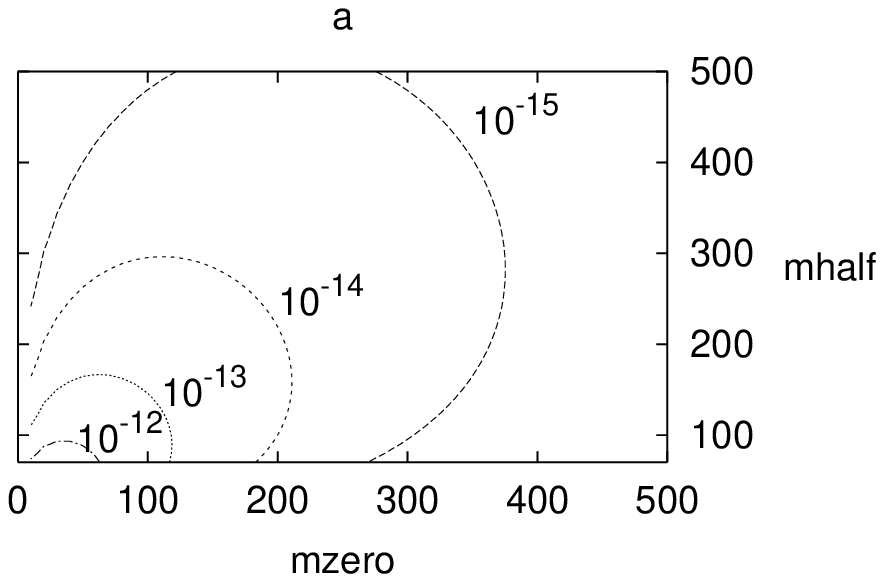}
        }
      }
    \caption{Contour plots for Br$(\mu \to e \gamma)$ 
      in the $m_0$ -- $M_{1/2}$ plane for $(a)~\rho_R = 1, ~a_1^R = 0$ and 
      $(b)~\rho_R = 0.1,~a_1^R = m_0$.
      The lines correspond to Br$(\mu \to e \gamma) = 10^{-12}, 10^{-13}, 10^{-14}$ 
      and $10^{-15}$ from the lower left.
      Here, we fix the $S_3$ breaking parameters at 
      $\delta_l =  6.577 \times10^{-2}$ 
      and $\epsilon_l = 4.792 \times 10^{-3}$ with 
      $y^l_0 \langle H \rangle = 2456 ~\mbox{MeV}$ and $r_R = 1$.}
    \label{fig:contour}
  \end{center}
\end{figure}

Next we would like to present some numerical results. 
We calculate the rate of the $\mu\to e\gamma$ by using the
formulae presented in Appendix.
For simplicity, we take a
universal gaugino mass $M_{1/2}$:
\begin{eqnarray}
  M_{\widetilde{B}}(M_X) = M_{\widetilde{W}}(M_X) = M_{\widetilde{G}}(M_X) 
  = M_{1/2},
\end{eqnarray}
where $M_{\widetilde{B}}, M_{\widetilde{W}}$ and $M_{\widetilde{G}}$ represent
bino, wino and gluino masses respectively,
and assume the equality of the right-handed slepton masses 
in the first two generations and the left-handed 
slepton masses $m^2_{R0}(M_X) = m^2_{L0}(M_X) = m^2_0$ 
at the boundary $M_X = 2 \times 10^{16}$ GeV.
The masses at low energy are obtained with the help of renormalization 
group equations. Relevant quantities in our analysis are
\begin{eqnarray}
  M_{\widetilde{B}}(\mu) & = & 0.43 M_{1/2}, \quad M_{\widetilde{W}}(\mu) = 0.83 M_{1/2}, \nonumber \\
  (Y_l)_{ij}(\mu) &=& 1.5\,(Y_l)_{ij}(M_X)\,, \nonumber 
  \\
  \left(m_{LL}^2\right)_{ij}(\mu) &=& \left(m_{LL}^2\right)_{ij}(M_X)
  + 0.50\,|M_{1/2}|^2 \delta_{ij}\,, \nonumber 
  \\
  \left(m_{RR}^2\right)_{ij}(\mu) &=& \left(m_{RR}^2\right)_{ij}(M_X)
  + 0.15\,|M_{1/2}|^2 \delta_{ij}\,,
  \\
  (A_l)_{ij}(\mu)&=& 1.5\,(A_l)_{ij}(M_X) + 0.67\,(Y_l)_{ij}(\mu) M_{1/2}\,,
\nonumber 
\end{eqnarray}
where the low-energy values are evaluated at the renormalization point
$\mu=500$ GeV, and we neglect the effect of the Yukawa couplings, which
is valid as far as the tau Yukawa couplings are small. For simplicity,
we fix the $S_3$ breaking parameters at $\delta_l =  6.577 \times10^{-2}$
and $\epsilon_l = 4.792 \times 10^{-3}$ with 
$y^l_0 \langle H \rangle = 2456 ~\mbox{MeV}$ and $r_R = 1$ 
(See eq. (\ref{eq:Delta-E})). Contours of the
branching ratio of $\mu \to e \gamma$ on the $m_0$ -- $M_{1/2}$ plane are
depicted in Fig. \ref{fig:contour}. 
Here we consider the two cases. In Fig. \ref{fig:contour} (a), 
we take $\rho_R=1$, $a_1^R=0$, and thus the $(m^2_{RR})_{12}$ is the
dominant source of lepton flavor violation. On the contrary, 
Fig. \ref{fig:contour} (b), 
is the case where $\rho_R=0.1$, $a_1^R=m_0$ so that the $(m^2_{RL})_{12}$
dominates over the other. We should compare the contours of the branching
ratio with the expected sensitivity of the branching ratio $\sim 10^{-14}$, 
which is aimed by the MEG collaboration at PSI \cite{meg}.  
The both figures show
that  $\mu \to e \gamma$ will be detectable if the  gaugino mass and
the scalar mass are smaller than about 300 GeV. We note that the
branching ratios are small when the universal scalar mass is small. This is
understood easily if we notice that the lepton flavor violation is inherent
in the scalar masses, and in this limit the effect is diluted by the 
large gaugino contributions coming from the renormalization group evolution.
We also checked other choices of $\rho_R$ and $a_1$. 
Generically, the results are not sensitive to the these values.
However, for a certain choice of the parameters, $a^R_1 \sim \rho_R m_0$,
an accidental cancellation between the two contributions to the
lepton flavor violation can take place,
and thus branching ratio will be significantly reduced.

In our analytic as well as numerical computations, we only considered the 
gaugino loops depicted as Fig. \ref{fig:lfv}. 
This is valid for small $\tan \beta$. 
In fact there are other diagrams which generically enhance the branching
ratio when $\tan \beta$ is large. Furthermore, it was pointed out in 
\cite{Hisano:1996qq}
that the contributions coming from various diagrams partially cancel with
each other in certain regions of the parameter space. This cancellation is,
however, very sensitive to the soft masses as well as the higgsino mass
parameter, and it would be premature to 
make a full analysis with various unknown parameters in the model.

Bearing these observations in mind, we conclude that $\mu \to e
\gamma$ is generically within the reach of the planning experiment in
a wide range of the parameter space if the masses of the
superparticles involved in the process are around the weak scale.

Here we would like to emphasize that 
considerable lepton flavor violation is expected in our scenario, even if 
the Yukawa couplings of the right-handed neutrinos are small and their
effects in the RG running are negligible. This clearly contrasts with 
other cases considered previously 
\cite{Borzumati:1986qx,Hisano:1995nq,Hisano:1997tc}, where 
the RG effect from the right-handed neutrino Yukawa couplings is essential 
for  sufficient lepton flavor violation.

Let us next consider other lepton flavor violating processes,
$\tau \to \mu \gamma$ and $\tau \to e \gamma$.
The ratios of the branching fractions are estimated as
\begin{eqnarray}
  \frac{\mbox{Br}(\tau \rightarrow \mu \gamma)}
  {\mbox{Br}(\mu \rightarrow e \gamma)}
  \sim 0.17 \left( \frac{\Delta_l}{\Delta_l E_l} \right)^2 \sim 10^3, \quad
  \frac{\mbox{Br}(\tau \rightarrow e \gamma)}
  {\mbox{Br}(\mu \rightarrow e \gamma)}
  \sim 0.17 \left( \frac{E_l}{\Delta_l E_l} \right)^2 \sim 10
\end{eqnarray}
where the factor $0.17$ is the branching ratio of $\tau \to \mu \nu_{\tau}
\bar{\nu}_{\mu} ~(e \nu_{\tau} \bar{\nu}_e)$.   
Recalling the fact that the current
experimental bound of $\mbox{Br}(\mu \to e \gamma)$ is already $\sim
10^{-11}$, we find that $\mbox{Br}(\tau \rightarrow \mu \gamma)$ expected 
in our scenario can
be at best $10^{-8}$, roughly two orders of magnitudes lower than its
present experimental bound \cite{Ahmed:1999gh},
and it may also be challenging for a Super B factory,
which is to explore down to $10^{-7} - 10^{-8}$ \cite{tmg}.

Finally, we would like to mention   CP phases from  the slepton sector.
Complex phases in the slepton masses generally induce CP violation. 
The strongest constraint on the complex phases comes from 
the electron electric dipole moment (EDM) \cite{Regan:ta}:
\begin{eqnarray}
  \left| \mbox{Im} (\delta^l_{11})_{LR} \right| 
  <  ({\rm a~few}) \times 10^{-8} 
  \left( \frac{m_{\tilde{l}}}{100 ~\mbox{GeV}} \right)^2.
\end{eqnarray}
As is evident in
eq.~(\ref{eq:A-sleptons}), $(m^{l2}_{11})_{LR}$ has two terms in our
scenario: the first term is a universal one, and the second term is
flavor-dependent term which originates from the democratic part of the
K\"ahler potential. We find that the former contribution dominates,
whose magnitude is estimated as
\begin{eqnarray}
  (\delta^l_{11})_{LR}
  \sim m_e a_0 / m_{\tilde{l}}^2 \sim 5 \times 10^{-6}.
\end{eqnarray}
To survive the experimental constraint, the phase should be rather small
as $\sim 0.1-0.01$. We note that this universal contribution exists in many
types of SUSY breaking scenario, including the minimal supergravity model, and
the suppression of the phase of the universal $A$-term is a common problem in
supersymmetry.\footnote{
The CP phase of the Higgs mixing mass term should also be suppressed.
See a recent proposal on this issue \cite{Yamaguchi:2002zy}.}
 Although our scenario does not solve this problem, it does not
worsen the situation either.

\section{FCNC  in Quark Sector}
Finally, we would like to briefly discuss FCNC constraints in the
quark sector in a  similar manner as
 in the previous
section.  Let us  consider the $K,B$ and $D$ mass differences
($\Delta m_K, \Delta m_B$ and $\Delta m_D$) and the branching fraction of the 
$b \to s \gamma$ process.
They provide the upper bounds of off-diagonal
elements of squark masses, which are listed in Table \ref{tab:fcnc}.
Here we quote the estimates in Ref. \cite{Gabbiani:1996hi}, 
which considers only gluino exchange diagrams,
and
all squark and gluino masses are fixed at
$m_{\tilde{q}} = 500$ GeV for simplicity. 
The limits are scaled as 
$(m_{\tilde{q}}/ 500 ~\mbox{GeV})$ for $\Delta m_K, \Delta m_B$ and 
$\Delta m_D$ and $( m_{\tilde{q}} / 500 ~\mbox{GeV} )^2$  
for $b \to s \gamma$.
The magnitudes of the off-diagonal elements expected  in this model 
are also listed in Table \ref{tab:fcnc}.
When  compared with the experimental bounds,
we conclude that
FCNC processes induced by the $S_3$ symmetry breaking effect
reside within the experimental bounds.

As for $b \to s \gamma$, there are other dangerous contributions from
a chargino-stop loop and a charged Higgs loop. 
It is known
that these two can contribute destructively for a sign of the higgsino
mixing parameter $\mu$, and thus in certain regions of the parameter
space the whole SUSY contribution is in accord with the experimental
constraint. We expect that this occurs and will not discuss this point
further.

The  CP violating quantities
$\epsilon_K, \epsilon'/\epsilon$ of the kaon system put severer constraints
on the imaginary parts of the flavor mixing masses.
One finds that the constraint from the $\epsilon_K$ bound to the product 
$(\delta^d_{12})_{LL}(\delta^d_{12})_{RR}$ is much stronger than
those to $(\delta^d_{12})_{LL}$ or $(\delta^d_{12})_{RR}$:
\begin{eqnarray}
  \sqrt{|\mbox{Im} (\delta^d_{12})_{LL}(\delta^d_{12})_{RR}|} 
  < 2.2 \times 10^{-4} \left( \frac{m_{\tilde{q}}}{500 ~\mbox{GeV}} \right) .
\end{eqnarray}
On the other hand, the magnitude of this combination is generically 
expected to be
\begin{eqnarray}
  \sqrt{(\delta^d_{12})_{LL}(\delta^d_{12})_{RR}}
  \sim \Delta_d E_d \sim 10^{-3},
\end{eqnarray}
one order of magnitude above the experimental constraint, which may
require some fine tuning to suppress the complex phase. Here, however,
we argue that this constraint is easily overcome if the model is SU(5)
inspired. In fact, if the model is embedded into SU(5), the
right-handed down-type squark mass matrix is forced to become an almost
unit matrix as well as the left-handed slepton one. This is simply
because the two are in $5^*$ representation in the SU(5) and the
left-handed sleptons are argued to have essentially no flavor
violation, which is suggested by the bi-large neutrino mixing
angles. Then $ (\delta^d_{12})_{RR}$ is  zero, and hence the
gluino contribution to $\epsilon_K$ vanishes.  The upper
bounds from the experiments and the expectations in our model for
other processes are listed in Table \ref{tab:cp} for the SU(5) case.
From this table, one finds that our scenario marginally survives the
bounds from  the CP violating quantities.

\section{Conclusions and discussion}

In this paper, we have proposed the extension of the democratic
ansatz to the sector of superpartners of fermions (sfermions) in the
Standard Model. It has been shown that the sfermion masses and their
mixing controlled by the democratic ansatz solves the SUSY flavor
problem and provides some phenomenologically interesting consequences.

More concretely we have postulated that the K\"ahler potential for the
chiral supermultiplets of the quarks and leptons possesses the
underlying $S_3$ symmetries of the democratic principle while the
superpotential contains their small violation to obtain realistic
Yukawa interactions for the quarks and leptons. Furthermore we have
assumed that the hidden sector which is responsible to the SUSY
breaking couples to the MSSM sector only in the K\"ahler potential,
which is actually a usual assumption in the hidden sector SUSY
breaking scenario. 

In this framework, we have shown that the expectations from the $S_3$
symmetries and their violation are 1) the masses of the third
generation sfermions can be different from those of the corresponding
first two generation sfermions, and 2) the small violation of the
$S_3$ symmetry induces the flavor violation in the sfermion
sector. Remarkably the requirement that the neutrino sector should
have bi-large mixing angles constrains the form of the K\"ahler metric
of the left-handed lepton doublets. Unlike the general expectations
given above, the requirement implies that the left-handed sleptons
should have degenerate masses among the different generations, and
also no flavor violation should arise in the left-handed slepton
mass matrix. 

This observation brings interesting phenomenological implications. The
slepton mass spectrum should be of a special form. The left-handed sleptons
have degenerate masses, while in the right-handed slepton sector the third
generation will have a different mass  from the first two generations.
These masses are given at some high energy scale and suffer from radiative
corrections. We expect that renormalization group analysis will relate the 
masses which will be measured experimentally with the masses given here.
Thus we hope that the mass measurement of the slepton masses which is
expected to be done in future collider experiments such as an $e^+ e^-$ 
linear collider will provide a test of the democratic sfermion ansatz.

A special attention has been paid to the lepton flavor violating
process $\mu \to e \gamma$. We have shown that the branching ratio
expected in our scenario  survives the current experimental bound,
and more interestingly it is generically within the reach of the
forthcoming experiment \cite{meg} whose goal is to explore the branching ratio
down to $\sim 10^{-14}$, if the sleptons and the charginos/neutralinos
have masses around the weak scale. Furthermore, in our scenario, the
lepton flavor violating operator of the dipole moment type appears only for
the right-handed anti-muon, and thus this property can be tested if the
polarized muon beam is available.  On the other hand, the lepton
flavor violation of the $\tau$ decay may not be a very promising process
to see a signal of our scenario.

We should emphasize that the origin of the lepton flavor violation in
the democratic sfermion ansatz is in the non-universality of the
sfermion masses allowed by the democratic principle. This should be
compared with the other mechanism to generate the lepton flavor
violation that the renormalization group effects from the right-handed
neutrino Yukawa interactions or the Yukawa interaction above the GUT
scale induces the lepton flavor mixing in the slepton mass matrix which
is assumed to be universal at the scale where these masses are
originally generated.

\section*{Acknowledgment}               
This work was supported in part by the
Grant-in-aid from the Ministry of Education, Culture, Sports, Science
and Technology, Japan,  No.12047201. KH thanks the Japan Society for
the Promotion of Science for financial support.

\appendix

\section{LFV decay rates and electron EDM}
In this appendix, we calculate
lepton flavor violating decay rates and 
the EDM of the electron, which arise from dipole moment operators.

We calculate the dipole moment operators using the mass insertion method.
Here we discard higgsino-wino mixing diagrams ($\mu$-term contribution)
and double mass insertion diagrams (contributions from e.g.,
$(m_{LL}^2)_{ij}\times (m_{LR}^2)_{jj}$), which are proportional to 
$\mu \tan \beta$. Thus expressions below are valid in low $\tan \beta$ region.
The resulting dipole moment operators are given by
\begin{eqnarray}
    {\cal L}_{\rm eff}  &= & e \frac{m_{l_j}}{2} \bar{l}_i \sigma_{\mu \nu} 
  F^{\mu \nu} (L_{ij} P_L + R_{ij} P_R) l_j
\end{eqnarray}
where
\begin{eqnarray}
  L_{ij} &=&
  \frac{\alpha}{2 \pi \cos^2 \theta_W}
  \frac{\left(m_{RR}^2\right)_{ij}}{m_{{\tilde{l}_R}}^4}
  M(x_R^B)
  -
  \frac{1}{2} \frac{\alpha}{2 \pi \cos^2 \theta_W}
  \left(\frac{M_{\widetilde{B}}}{m_{lj}}\right)
  \frac{\left(m_{RL}^2\right)_{ij}}{m_{{\tilde{l}_L}}^2 m_{{\tilde{l}_R}}^2}
  \frac{T(x_L^B) - T(x_R^B)}{x_L^B - x_R^B}
  \,, \nonumber \\
  R_{ij} &=&
  \frac{1}{4} \frac{\alpha}{2 \pi \sin^2 \theta_W}
  \frac{\left(m_{LL}^2\right)_{ij}}{m_{\tilde{l}_L}^4}
  \left[
    \tan^2\theta_W M(x_L^B)
    + M(x_L^W)
    + S(x_L^W)
  \right]
  \nonumber\\
  &&
  -
  \frac{1}{2} \frac{\alpha}{2 \pi \cos^2 \theta_W}
  \left(\frac{M_{\widetilde{B}}}{m_{lj}}\right)
  \frac{\left(m_{LR}^2\right)_{ij}}{m_{\tilde{l}_L}^2 m_{\tilde{l}_R}^2}
  \frac{T(x_L^B) - T(x_R^B)}{x_L^B - x_R^B}
  \,.
\end{eqnarray}
Here $\alpha$ is the fine-structure constant, $\sin\theta_W$ is
the weak-mixing angle, $G_F$ is the Fermi constant.
$m_{{\tilde{l}}_{L,R}}$ represent the average of slepton masses of 
left- and right-handed types.
We define
\begin{eqnarray}
  &&
  x_L^B
  \equiv
  \frac{M_{\widetilde{B}}^2}{m_{{\tilde{l}_L}}^2}
  \,,\quad
  x_R^B
  \equiv
  \frac{M_{\widetilde{B}}^2}{m_{{\tilde{l}_R}}^2}
  \,,\quad
  x_L^W
  \equiv
  \frac{M_{\widetilde{W}}^2}{m_{{\tilde{l}_L}}^2}
  \,,
\end{eqnarray}
and the functions of $x$:
\begin{eqnarray}
  M(x)&\equiv&
  \frac{1-9x-9x^2+17x^3-6(3+x)x^2\log x}{12(1-x)^5}\,, \nonumber \\
  S(x)&\equiv&
  \frac{-1-9x+9x^2+x^3-6(1+x)x\log x}{3(1-x)^5}\,, \nonumber \\
  T(x)&\equiv&
  \frac{-x+x^3-2x^2\log x}{2(x-1)^3}\,.
\end{eqnarray}
From the above expressions, we obtain
the branching ratios for the processes $l_j \to l_i \gamma$, which are given by
\begin{eqnarray}
  {\rm Br}(l_j\to l_i\gamma) 
  &=&  
  \frac{48\pi^3 \alpha}{G_F^2}
  \left(
    \left|L_{ij}\right|^2
    +
    \left|R_{ij}\right|^2
  \right)
  \times
  {\rm Br}(l_j\to l_i \bar{\nu}_i \nu_j),
\end{eqnarray}
and the EDM of the electron by
\begin{eqnarray}
  \frac{d_e}{e} 
  = - m_e \, \mbox{Im} (R_{11})
  = 
  \frac{\alpha M_{\widetilde{B}}}{4 \pi \cos^2 \theta_W}
  \frac{\mbox{Im} [(m_{LR}^2)_{11}]}{m_{{\tilde{l}_L}}^2 m_{{\tilde{l}_R}}^2}
  \frac{T(x_L^B) - T(x_R^B)}{x_L^B - x_R^B}.
\end{eqnarray}


\begin{figure}[h]
  \begin{center}
    \begin{picture}(400,100)(0,0) 
      \Line(20,20)(60,20) \DashLine(60,20)(140,20){5} \Line(140,20)(180,20)
      \Text(40,10)[]{$l_j$} \Text(80,10)[]{$\tilde{l}_{Xj}$}
      \Text(120,10)[]{$\tilde{l}_{Xi}$} \Text(160,10)[l]{$l_i$}
      \Vertex(40,20){3} \Text(40,30)[]{$m_{lj}$}
      \Vertex(100,20){3} \Text(100,30)[]{$(m^2_{XX})_{ij}$}
      \PhotonArc(100,20)(40,0,180){3}{10} \CArc(100,20)(40,0,180)
      \Text(100,70)[]{$\widetilde{B}~(\widetilde{W})$}
      \Photon(140,60)(170,90){3}{6} \Text(170,70)[]{$\gamma$}
      \Line(220,20)(260,20) \DashLine(260,20)(340,20){5} \Line(340,20)(380,20)
      \Text(240,10)[]{$l_j$} \Text(280,10)[]{$\tilde{l}_{Yj}$}
      \Text(320,10)[]{$\tilde{l}_{Xi}$} \Text(360,10)[l]{$l_i$}
      \Vertex(300,20){3} \Text(300,30)[]{$(m^2_{XY})_{ij}$}
      \PhotonArc(300,20)(40,0,180){3}{10} \CArc(300,20)(40,0,180)
      \Text(300,70)[]{$\widetilde{B}$}
      \Vertex(300,60){3} \Text(300,50)[]{$M_{\widetilde{B}}$}
      \Photon(340,60)(370,90){3}{6} \Text(370,70)[]{$\gamma$}
    \end{picture}
    \caption{Feynman diagrams which cause lepton flavor violating decays
      $\l_j \to l_i \gamma$. 
      $l_i, \tilde{l}_{Xi}, \gamma, \widetilde{W}$ and $\widetilde{B}$ 
      represent the charged leptons,
      charged sleptons, the photon, wino and bino, respectively.
      $m_{lj}$ and $M_{\widetilde{B}}$ are the masses of 
      the charged leptons and bino.
      The blobs in the slepton lines indicate the insertions of 
      the flavor violating mass matrix elements, 
      $(m^2_{XX})_{ij}$ and $(m^2_{XY})_{ij}$,
      and ones in the fermion lines indicate chirality flips of
      the charged leptons and bino.
      $X,Y$ stand for the chiralities $L$ or $R$.}
    \label{fig:lfv}
  \end{center}
\end{figure}
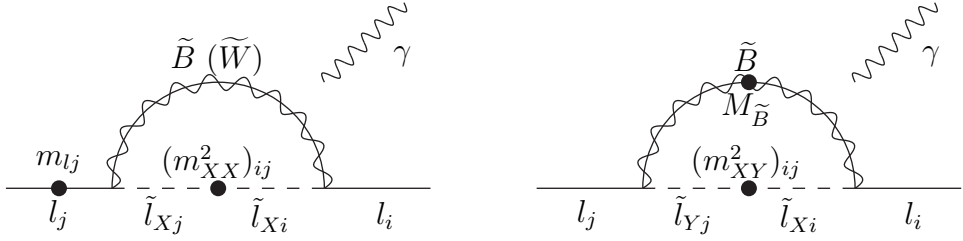

\begin{table}[htbp]
  \begin{center}
    \caption{Limits on $\delta$'s obtained from the FCNC experiments
      and estimation in our model.
      We fix all squark masses $m_{\tilde{q}}$, gluino masses and 
      $a_1$'s at $500$ GeV as representative values.
      For other values of $m_{\tilde{q}}$,
      the limits can be obtained multiplying the ones in the table by 
      $(m_{\tilde{q}}/ 500 ~\mbox{GeV})$ for $\Delta m_K, \Delta m_B$ and 
      $\Delta m_D$ and $( m_{\tilde{q}} / 500 ~\mbox{GeV} )^2$  
      for $b \to s \gamma$. 
      Notice that $(\delta_{ij})_{LR} \sim (\delta_{ij})_{RL}$ are imposed in 
      obtaining the limits. It does not matter since we focus on 
      order of magnitudes.}
    \label{tab:fcnc}
    \begin{eqnarray*}
      \begin{array}{ccccc}
        & {\rm Exp.~ bounds}\cite{Gabbiani:1996hi} & {\rm Our~model} \\ 
        \Delta m_K \\
        \sqrt{|\mbox{Re} (\delta^d_{12})_{LL,RR}^2|} & 4.0 \times 10^{-2} & 
        \Delta_d E_d \sim 10^{-3} \\
        \sqrt{|\mbox{Re} (\delta^d_{12})_{LR,RL}^2|} & 4.4 \times 10^{-3} & 
        a_1^{L,R} \Delta_d E_d m_s/ m_{\tilde{d}}^2 \sim 10^{-6} \\
        \sqrt{|\mbox{Re} (\delta^d_{12})_{LL}(\delta^d_{12})_{RR}|} & 2.8 \times 10^{-3} & 
        \Delta_d E_d \sim 10^{-3}  \\ \\
        \Delta m_B \\
        \sqrt{|\mbox{Re} (\delta^d_{13})_{LL,RR}^2|} & 9.8 \times 10^{-2} & 
        E_d \sim 10^{-2} \\
        \sqrt{|\mbox{Re} (\delta^d_{13})_{LR,RL}^2|} & 3.3 \times 10^{-2} & 
        a_1^{L,R} E_d m_b/ m_{\tilde{d}}^2 \sim 10^{-4} \\
        \sqrt{|\mbox{Re} (\delta^d_{13})_{LL}(\delta^d_{13})_{RR}|} & 1.8 \times 10^{-2} & 
        E_d \sim 10^{-2} \\ \\
        \Delta m_D \\
        \sqrt{|\mbox{Re} (\delta^u_{12})_{LL,RR}^2|} & 0.10 & 
        \Delta_u E_u \sim 10^{-6} \\
        \sqrt{|\mbox{Re} (\delta^u_{12})_{LR,RL}^2|} & 3.1 \times 10^{-2} & 
        a_1^{L,R} \Delta_u E_u m_c/m^2_{\tilde{u}} \sim 10^{-8} \\
        \sqrt{|\mbox{Re} (\delta^u_{12})_{LL}(\delta^u_{12})_{RR}|} & 1.7 \times 10^{-2} & 
        \Delta_u E_u \sim 10^{-6} \\ \\
        b \to s \gamma \\
        |(\delta^d_{23})_{LL,RR}| & 8.2 & 
        \Delta_d \sim 10^{-1} \\
        |(\delta^d_{23})_{LR,RL}| & 1.6 \times 10^{-2} & 
        a_1^{L,R} \Delta_d m_b/m^2_{\tilde{u}} \sim 10^{-3}
      \end{array}
    \end{eqnarray*} 
  \end{center}
\end{table}

\begin{table}[htbp]
  \begin{center}
    \caption{Limits on $\delta$'s obtained from the CP violating parameters
      and estimation in our model in the SU(5) GUT inspired case.
      We use the same parameters as in Table I.
      For other values of $m_{\tilde{q}}$,
      the limits can be obtained multiplying the ones in the table by 
      $(m_{\tilde{q}}/ 500 ~\mbox{GeV})$ for $\epsilon_K$
      and $( m_{\tilde{q}} / 500 ~\mbox{GeV} )^2$  
      for $\epsilon'/\epsilon$.}
    \label{tab:cp}
       \begin{eqnarray*}
      \begin{array}{ccccc}
        & {\rm Exp.~ bounds}\cite{Gabbiani:1996hi} & {\rm Our~ model} & \\ 
        \epsilon_K \\
        \sqrt{|\mbox{Im} (\delta^d_{12})_{LL}^2|} & 3.2 \times 10^{-3} & 
        \Delta_d E_d \sim 10^{-3} \\
        \sqrt{|\mbox{Im} (\delta^d_{12})_{RR}^2|} & 3.2 \times 10^{-3} & 
        \sim 0 \\
        \sqrt{|\mbox{Im} (\delta^d_{12})_{LR}^2|} & 3.5 \times 10^{-4} & 
        a_1^L \Delta_d E_d m_s/m^2_{\tilde{d}} \sim 10^{-6} \\
        \sqrt{|\mbox{Im} (\delta^d_{12})_{RL}^2|} & 3.5 \times 10^{-4} & 
        a_1^L \Delta_d E_d m_d/m^2_{\tilde{d}} \sim 10^{-8} \\
        \sqrt{|\mbox{Im} (\delta^d_{12})_{LL}(\delta^d_{12})_{RR}|} & 2.2 \times 10^{-4} & 
        \sim 0 \\ \\
        \epsilon'/\epsilon \\
        |\mbox{Im} (\delta^d_{12})_{LL}| & 0.48 & 
        \Delta_l E_l \sim 10^{-3} \\
        |\mbox{Im} (\delta^d_{12})_{RR}| & 0.48 & 
        \sim 0 \\
        |\mbox{Im} (\delta^d_{12})_{LR}| & 2.0 \times 10^{-5} & 
        a_1^L \Delta_d E_d m_s/m^2_{\tilde{l}} \sim 10^{-6} \\
        |\mbox{Im} (\delta^d_{12})_{RL}| & 2.0 \times 10^{-5} & 
        a_1^L \Delta_l E_l m_d/m^2_{\tilde{l}} \sim 10^{-8}
      \end{array}
    \end{eqnarray*}
  \end{center}
\end{table}

\end{document}